\documentclass[twocolumn,showpacs,preprintnumbers,amsmath,amssymb, showkeys]{revtex4}

\usepackage{graphicx}
\usepackage{dcolumn}
\usepackage{bm}
\usepackage{epsf}

\begin{document}


\title{Localized charge bifurcation in the coupled quantum dots}

\author{V.\,N.\,Mantsevich}
 \altaffiliation{vmantsev@spmlab.phys.msu.ru}
\author{N.\,S.\,Maslova}%
 \email{spm@spmlab.phys.msu.ru}
\author{P.\,I.\,Arseyev}%
 \email{ars@lpi.ru}

\affiliation{%
 Moscow State University, Department of  Physics,
119991 Moscow, Russia
}%
\affiliation{%
 P.N. Lebedev Physical institute of RAS, Moscow, Russia, 119991
}%

\date{\today }

\begin{abstract}
We theoretically analyzed localized charge relaxation in a double
quantum dot (QD) system coupled with continuous spectrum states in
the presence of Coulomb interaction between electrons within a dot.
We have found that for a wide range of the system parameters charge
relaxation occurs through two stable regimes with significantly
different relaxation rates. A certain instant of time exists in the
system at which rapid switching between stable regimes takes place.
We consider this phenomenon to be applicable for creation of active
elements in nano-electronics based on the fast transition effect
between two stable states.
\end{abstract}

\maketitle

\section{Introduction}

 Nano-scale electronics is currently a very active area of
research. One of the main goals in this field is to design and
characterize low dimensional structures that could be active
elements in electronic circuitry \cite{Collier},\cite{Gittins}.
Single semiconductor QDs which are referred as "artificial" atoms
\cite{Kastner},\cite{Ashoori} and coupled QDs - "artificial"
molecules \cite{Oosterkamp},\cite{Blick_0} are usually suggested as
perspective structures that may serve for creation of extremely
small devices because of the possibility for the electrons spatial
confinement on a scale less than 10 nm due to the growth processes
\cite{Bimberg}. Double QDs systems behaviour intrinsically differs
from a single QDs because of the variable interdot tunneling
coupling \cite{Oosterkamp},\cite{Livermore}, which is the physical
reason for non-linearity formation and consequently for existence of
such phenomena as bifurcations \cite{Rotter} and bistability
\cite{Goldman}. That's why double QDs can be applied for logic gates
fabrication based on the effect of ultrafast switching between
intrinsic stable states. During the last decade vertically aligned
QDs have been fabricated and widely studied with the great success
(for example indium arsenide QDs in gallium
arsenide)\cite{Vamivakas},\cite{Stinaff},\cite{Elzerman}. The
applied gate voltage dictates the electron occupancy of each QD via
a nearby electron reservoir and tunes the relative energy separation
between the electronic states of the two QDs \cite{Vamivakas}. It
was demonstrated experimentally that the localized states with
different charge and spin configuration can be tuned into the
resonance with external optical field due to the presence of Coulomb
interaction within QDs \cite{Stinaff}. These effects can lead to
inverse occupation of different localized states, so, fully
controllable solid-state single-emitter laser can be produced on the
system of coupled QDs \cite{Elzerman}. Lateral QDs seems to be a
betters candidate for scaling up the electronic coupling from two or
several QDs by applying individual lateral gates. That's why they
are intensively studied in the last several years both
experimentally and theoretically \cite{Peng},\cite{Munoz-Matutano}.
Nano-devices that exhibit fast switching are supposed to be a basis
of future oscillators, amplifiers and other important circuit
elements. The technological problem of QDs integration in a little
quantum circuits deals with the careful analysis of non-equilibrium
charge distribution, relaxation processes and non-stationary effects
influence on the electron transport through the system of QDs
\cite{Angus},\cite{Grove-Rasmussen},\cite{Moriyama},\cite{Landauer}.
Electron transport in such systems is strongly governed by the
presence of Coulomb interaction between electrons within a dot and
of course by the ratio between the QDs coupling and coupling to the
leads \cite{Mantsevich}. So the problem of charge relaxation due to
the tunneling between QDs coupled with continuous spectrum states in
the presence of Coulomb interaction is really vital.

Intrinsic bistabilities in different tunneling structures were
widely studied experimentally and theoretically. Obtained results
provide evidence for various molecular \cite{Collier},\cite{Gittins}
and QDs \cite{Alexandrov},\cite{Orellana},\cite{Rack},\cite{Djuric}
switching effects apparent in I-V characteristics due to the
bistabilities in the tunneling current flowing through the system.
It was experimentally demonstrated \cite{Collier},\cite{Gittins}
that switching strongly depends on the choice of contacts,
substrates and can be observed even for simple molecules. Coulomb
interaction in such systems can be a reason for transitions between
stable states. The role of Coulomb interaction in double QDs
bistability formation was experimentally investigated in
\cite{Goldman}. Authors revealed that double-barrier
resonant-tunneling structures have intrinsic bistable behaviour in
I-V characteristics due to nonlinearities introduced by the Coulomb
interaction and demonstrate two branches with high and low current
for the same voltage. Theoretical investigations of bistable
behaviour in tunneling structures usually deal with slave-boson
technique \cite{Orellana},\cite{Coleman_1},\cite{Coleman_2},
drift-diffusion approach \cite{Rack},\cite{Wetzler} or Hubbard
approximation \cite{Hubbard} with negative values of Coulomb
interaction $U$ \cite{Alexandrov}.

The further progress in electronics will depend upon understanding
intrinsic mechanisms for molecules and coupled QDs reversible
switching from low to high current states but this question is still
not well understood. Moreover this effect can't be accurately
controlled. That's why we consider bifurcations in the system of QDs
to be much more suitable mechanism for fast switching circuits
creation. Bifurcation means that a system has several stable states
or evolution regimes separated in time. Conditions for ultrafast
switching between these states can be controled by means of changing
energy levels positions in QDs, the value of Coulomb interaction and
strength of QDs coupling.

In this paper we consider charge relaxation within coupled QDs due
to the tunneling to continuous spectrum states in the presence of
Coulomb interaction between electrons within a QD by means of
Keldysh diagram technique \cite{Keldysh}. Tunneling to the continuum
is possible only from one of the QDs. We have found that for a wide
range of the system parameters charge relaxation occurs through two
stable regimes with different relaxation rates. At a certain instant
of time system switches rapidly between the regimes.

\section{Theoretical model}

\begin{figure} [t]
\includegraphics{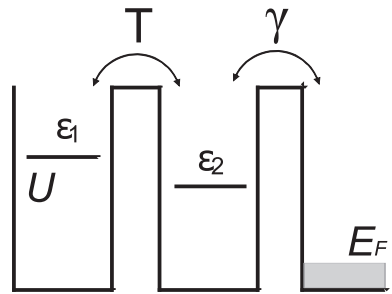}
\caption{Schematic diagram of energy levels in the system of two
coupled QDs. Second QD is also connected with continuous spectrum
states.}\label{figure1}
\end{figure}

The model under investigation deals with the system of two coupled
QDs with energy levels $\varepsilon_1$ and $\varepsilon_2$
correspondingly (Fig.\ref{figure1}). QD with energy level
$\varepsilon_2$ is also connected with the continuous spectrum
states. Hamiltonian of the system can be written as:
\begin{eqnarray}
\hat{H}=\varepsilon_{1}c_{1}^{+}c_{1}+\varepsilon_{2}c_{2}^{+}c_{2}+\sum_{k}\varepsilon_{k}c_{k}^{+}c_{k}+\nonumber\\
+T(c_{1}^{+}c_{2}+c_{2}^{+}c_{1})+\sum_{k}T_{k}(c_{k}^{+}c_{2}+c_{2}^{+}c_{k})
\end{eqnarray}

where $T$ and $T_{k}$ are tunneling transfer amplitudes between the
QDs and between the second QD and continuous spectrum states
correspondingly. By considering the constant density of state
$\nu_{k}^{0}$ in the continuous spectrum the tunneling relaxation
rate $\gamma$ is defined as $\gamma=\pi\nu_{k}^{0}T_{k}^{2}$.
$c_{1}^{+}/c_{1}$($c_{2}^{+}/c_{2}$) and $c_{k}^{+}/c_{k}$-
electrons creation/annihilation operators in the first(second) QD
localized state and in the continuous spectrum states ($k$)
correspondingly. We also take into account on-site Coulomb repulsion
in the quantum dot with energy level $\varepsilon_1$ (first QD).
Interaction Hamiltonian has the form:

\begin{eqnarray}
H_{int}=Un_{1\sigma}n_{1-\sigma}
\end{eqnarray}

Different relaxation regimes are determined by the relations between
the model parameters: $T$, $\gamma$, $U$ and $\varepsilon_i$. QDs
electronic states coupling $T$ is determined by the distance between
the QDs, the barrier height and localized states energy levels
positions. Energy levels positions are strongly connected with the
QDs geometry: the width and the depth of potential well associated
with each dot. The Coulomb interaction between localized electrons
is governed by the localization radius of electronic states within
the QDs. The value of tunneling rate $\gamma$ depends on the width
and height of the barrier which separates second QD from the lead.
So all these parameters can be varied in the real experimental
situation. Our model parameters correspond to the experimental
situation when one of the vertically stacked interacting QDs is deep
and narrow (deep energy levels and Coulomb interaction must be taken
into account) and another one is wide and shallow (shallow energy
levels and Coulomb interaction is small and can be
neglected)\cite{Kikoin}.

As we are interested in the specific features of the non-stationary
time evolution of the initially localized charge within the coupled
QDs, we'll consider the situation when condition
$(\varepsilon_i-\varepsilon_F)/\gamma)>>1$ is fulfilled. It means
that initial energy levels are situated well above the Fermi level
and stationary occupation numbers in the second QD in the absence of
coupling between the QDs is of order
$\gamma/(\varepsilon_2-\varepsilon_F)<<1$.  So the Kondo effect
can't appear in the proposed model and we can also omit the terms
corresponding to the stationary solution.

First of all we shall find localized charge relaxation laws in the
coupled QDs without Coulomb interaction between localized electrons
in the first QD. Let us assume that at the initial moment all charge
density in the system is localized in the first QD and has the value
$n_{1}(0)$. In the absence of tunneling between the QDs Green
functions $G_{11}^{R}(t-t^{'})$ and $G_{22}^{R}(t-t^{'})$ can be
found from expressions:

\begin{eqnarray}
G_{11}^{R}(t-t^{'})&=&-i\Theta(t-t^{'})e^{-i\varepsilon_1(t-t^{'})}\nonumber\\
G_{22}^{R}(t-t^{'})&=&-i\Theta(t-t^{'})e^{-i\varepsilon_2(t-t^{'})-\gamma(t-t^{'})}
\end{eqnarray}

where $\gamma=\pi\nu_{k}^{0}T_{k}^{2}$ is tunneling relaxation rate
from the second QD to the continuous spectrum states.

Retarded electron Green's function $G_{11}^{R}$ determine spectrum
re-normalization due to tunneling processes between QDs and can be
found exactly from the integral equation:

\begin{eqnarray}
G_{11}^{R}=G_{11}^{0R}+G_{11}^{0R}T^{2}G_{22}^{R}G_{11}^{R}
\label{integral_equation}
\end{eqnarray}

Acting by inverse operators $G_{11}^{0R-1}$ and $G_{22}^{R-1}$
integral equation (\ref{integral_equation}) can be also presented in
the equivalent differential form (except the point $t=t^{'}$):

\begin{eqnarray}
((i\frac{\partial}{\partial
t}-\varepsilon_2+i\gamma)(i\frac{\partial}{\partial
t}-\varepsilon_1)-T^{2})G_{11}^{R}(t,t^{'})=0
\end{eqnarray}

Finally, retarded Green function $G_{11}^{R}$ can be written in the
following form:

\begin{eqnarray}
G_{11}^{R}(t,t^{'})=i\Theta(t-t^{'})(\frac{E_1-\varepsilon_2+i\gamma}{E_1-E_2}e^{-E_{1}(t-t^{'})}-\nonumber\\
-\frac{E_2-\varepsilon_2+i\gamma}{E_1-E_2}e^{-E_{2}(t-t^{'})})
\end{eqnarray}

where eigenfrequencies $E_{1,2}$ are determined by the equations:

\begin{eqnarray}
(E-\varepsilon_1)(E-\varepsilon_2+i\gamma)-T^{2}=0\nonumber\\
E_{1,2}=\frac{1}{2}(\varepsilon_1+\varepsilon_2-i\gamma)\pm\frac{1}{2}\sqrt{(\varepsilon_1-\varepsilon_2+i\gamma)^{2}+4T^{2}}
\label{E}
\end{eqnarray}

Let us now analyze time evolution of the electron density in the
considered system which is governed by the Keldysh Green function
$G_{11}^{<}(t,t^{'})$ \cite{Keldysh}:

\begin{eqnarray}
G_{11}^{<}(t,t^{'})=in_{1}(t)
\end{eqnarray}

Equation for Green function $G_{11}^{<}$ has the form:

\begin{eqnarray}
G_{11}^{<}(t,t^{'})=G_{11}^{0<}+G_{11}^{0<}T^{2}G_{22}^{A}G_{11}^{A}+\nonumber\\
+G_{11}^{0R}T^{2}G_{22}^{R}G_{11}^{<}+G_{11}^{0R}T^{2}G_{22}^{<}G_{11}^{A}\nonumber\\
\end{eqnarray}

and after acting by $G_{11}^{0R-1}$ can be re-written as:

\begin{eqnarray}
(G_{11}^{0R-1}-T^{2}G_{22}^{R})G_{11}^{<}=T^{2}G_{22}^{<}G_{11}^{A}
\label{equation}
\end{eqnarray}

Green function $G_{11}^{<}(t,t)$ is determined by the sum of
homogeneous and inhomogeneous solutions. Inhomogeneous solution of
the equation can be written in the following way:

\begin{eqnarray}
G_{11}^{<}(t,t^{'})=T^{2}\int_{0}^{t}dt_{1}\int_{0}^{t^{'}}dt_{2}G_{11}^{R}(t-t_{1})\cdot\nonumber\\
\cdot G_{22}^{<}(t_1-t_2)G_{11}^{A}(t_2-t^{'})\nonumber\\
\end{eqnarray}

If $G_{22}^{<}(0,0)=0$, Green function $G_{11}^{<}(t,t)$ is defined
by the solution of homogeneous equation. Homogeneous solution of the
differential equation has the form:

\begin{eqnarray}
G_{11}^{<}(t,t^{'})=f_{1}(t^{'})e^{-iE_{1}t}+f_{2}(t^{'})e^{-iE_{2}t}
\end{eqnarray}

It is also necessary to satisfy the symmetry relations for function
$G^{<}(t,t^{'})$:

\begin{eqnarray}
(G_{11}^{<}(t,t^{'}))^{*}=-G_{11}^{<}(t^{'},t)
\end{eqnarray}

We can determine all the coefficients because the solution has to
satisfy homogeneous integro-differential equation:

\begin{eqnarray}
G_{11}^{<}(t^{'},t)=iAe^{-iE_{1}t+iE_{1}^{*}t^{'}}+iBe^{-iE_{1}t+iE_{2}^{*}t^{'}}+\nonumber\\
+iB^{*}e^{-iE_{2}t+iE_{1}^{*}t^{'}}+iCe^{-iE_{2}t+iE_{2}^{*}t^{'}}
\end{eqnarray}

We also have to fulfill initial condition:

\begin{eqnarray}
G_{11}^{<}(0,0)=in_{1}^{0}
\end{eqnarray}

As far as solution has to satisfy homogeneous integro-differential
equation, after some calculations one can find the following
proportionality between $f_{1}(t^{'})$ and $f_{2}(t^{'})$:

\begin{eqnarray}
\frac{f_{1}(t^{'})}{f_{2}(t^{'})}=-\frac{\varepsilon_2-E_{1}-i\gamma}{\varepsilon_2-E_{2}-i\gamma}
\end{eqnarray}

Finally time dependence of filling numbers in the first QD
$n_{1}(t)$ can be written as:

\begin{eqnarray}
n_{1}(t)&=&n_{1}^{0}\cdot(Ae^{-i(E_{1}-E_{1}^{*})t}+2Re(Be^{-i(E_{1}-E_{2}^{*})t})+\nonumber\\
&+&Ce^{-i(E_{2}-E_{2}^{*})t}) \label{filling_numbers_1}
\end{eqnarray}

where coefficients $A$, $B$ and $C$ are determined as:

\begin{eqnarray}
A&=&\frac{|E_{2}-\varepsilon_1|^{2}}{|E_{2}-E_{1}|^{2}};
C=\frac{|E_{1}-\varepsilon_1|^{2}}{|E_{2}-E_{1}|^{2}}\nonumber\\
B&=&-\frac{(E_{2}-\varepsilon_1)(E_{1}^{*}-\varepsilon_1)}{|E_{2}-E_{1}|^{2}}
\label{p1}
\end{eqnarray}

Time evolution of electron density in the second QD is determined by
the Green function $G_{22}^{<}(t,t^{'})$ with initial condition
$G_{22}^{<}(0,0)=0$. Green function $G_{22}^{<}(t,t^{'})$ can be
found from equation similar to the equation (\ref{equation}) with
the following indexes changing ($1\leftrightarrow2$). Due to the
initial conditions $n_{2}(0)=0$, $n_{1}(0)=n_{0}=1$, filling numbers
evolution in the second QD $n_{2}(t)$ is defined by the
inhomogeneous part of the solution. So time dependence of the
electron filling numbers in the second QD $n_{2}(t)$ can be written
as:

\begin{eqnarray}
n_{2}(t)&=&(De^{-i(E_{1}-E_{1}^{*})t}+2Re(Ee^{-i(E_{1}-E_{2}^{*})t})+\nonumber\\
&+&Fe^{-i(E_{2}-E_{2}^{*})t}) \label{filling_numbers_3}
\end{eqnarray}

where coefficients $D$, $E$ and $F$ are determined by expressions:

\begin{eqnarray}
D=F=-E=\frac{T^{2}}{|E_{2}-E_{1}|^{2}} \label{p2}
\end{eqnarray}

It is clearly evident that three typical time scales exist in the
considered system in the absence of Coulomb interaction between
localized electrons, which are described by the expressions
(\ref{filling_numbers_1}),(\ref{filling_numbers_2}). Two of them we
shall identify as the first $|E_{1}-E_{1}^{*}|$ and second
$|E_{2}-E_{2}^{*}|$ mode correspondingly. One more time scale is
defined by the expression $|E_{1}-E_{2}^{*}|$ and results in
formation of charge density oscillations in both QDs, when the
following ratio between $T$ and $\gamma$ is valid:
$T/\gamma>1/\sqrt{2}$. Several time rates in localized charge
relaxation in a QD coupled with the thermostat were also found and
carefully analyzed in \cite{Contreras}.

It is necessary to mention that the suggested model can be
generalized for the situation when both QDs are connected with the
leads. All the expressions up to the equation (\ref{p2}) continue
being valid if one substitutes $\varepsilon_1$ by the expression
$\varepsilon_1-i\gamma_1$ where $\gamma_1$ is a tunneling rate from
the first QD to the contact lead. In this paper we are interested in
the specific features of charge relaxation processes first of all
due to the charge redistribution between the coupled QDs. So we
consider strongly asymmetric case when $\gamma_1<<\gamma$ and
$\gamma_1<<T$. In such approximation obtained results can be applied
to the system of coupled QDs connected with the both leads. The
presented assumptions correspond to the well known experimentally
vertically aligned geometry of the coupled QDs
\cite{Vamivakas},\cite{Stinaff},\cite{Elzerman}.

Now we shall take into account on-site Coulomb repulsion within the
first QD. We shall confine ourself by analyzing only paramagnetic
case when $n_{1\sigma}=n_{1-\sigma}=n_{1}$.

Coulomb interaction within the first QD is considered by means of
self-consistent mean field approximation \cite{Anderson}. It means
that in the final expressions for the filling numbers time evolution
(\ref{filling_numbers_1}),(\ref{filling_numbers_3}) one have to
substitute energy level value $\varepsilon_1$ by the value
$\widetilde{\varepsilon}_1$, which is determined as:

\begin{eqnarray}
\widetilde{\varepsilon}_1=\varepsilon_1+U+U\cdot(<n_1(t)>-1)
\label{Coulomb}
\end{eqnarray}

We consider charge relaxation from initially filled electronic state
in the first QD, and it is reasonable to determine initial energy
level position in the first QD as $\varepsilon_1+U$, where
$\varepsilon_1$ is the energy of the empty electronic state. So the
initial detuning is
$\Delta\varepsilon=\varepsilon_1+U-\varepsilon_2$ because at $t=0$:
$\widetilde{\varepsilon}_1=\varepsilon_1+U$.

Consequently one should solve self-consistent system of equations
(\ref{E}), (\ref{filling_numbers_1}), (\ref{p1}) and (\ref{Coulomb})
to obtain the new energy level position $\widetilde{\varepsilon}_1$
and $n_{1}(t)$. First of all it is necessary to substitute
expressions for $E_1$ and $E_2$ from eq.(\ref{E}) to the
eq.(\ref{p1}) and to determine coefficients $A$, $B$ and $C$. Than
one have to substitute $A$, $B$ and $C$ from eq.(\ref{p1}) to the
eq.(\ref{filling_numbers_1}). Finally two equations are obtained
where equation for $n_1(t)$ depends on the new energy level position
and equation for $\widetilde{\varepsilon}_1$ depends on the filling
numbers time evolution $n_1(t)$. These two equations can be solved
self-consistently. The result of self-consistent solution gives us
$n_1(t)$ and new energy level position as a functions of time. After
this procedure coefficients $D$, $E$, and $F$ can be found from eq.
(\ref{p2}). And finally substituting eq.(\ref{p2}) to
eq.(\ref{filling_numbers_3}) one can obtain $n_2(t)$.

\section{Calculation results}

\begin{figure*} [t]
\includegraphics[width=110mm]{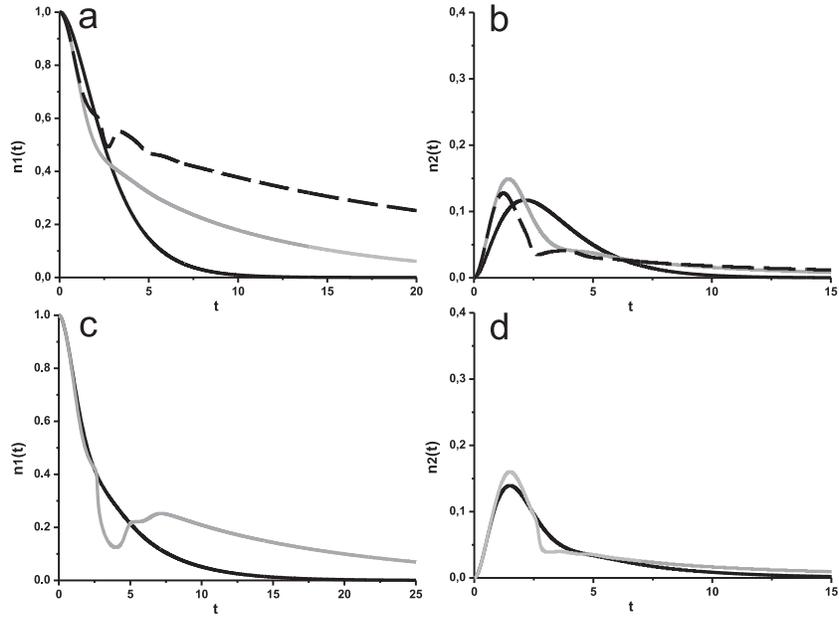}
\caption{Filling numbers evolution in the first a),c). $n_{1}(t)$
and second b),d). $n_{2}(t)$ QDs. Panels a). and b). correspond to
the resonant tunneling $(\varepsilon_1-\varepsilon_2)/\gamma=0.0$.
$U=0$-black line, $U=2$-grey line, $U=4$-dashed line. Panels c). and
d). correspond to the tunneling between QDs in the presence of
detuning $(\varepsilon_1=\varepsilon_2)/\gamma=1.0$. $U=0$-black
line, $U=2$-grey line. Parameters $T/\gamma=0.6$ and $\gamma=1.0$
are the same for all the figures.} \label{figure2}
\end{figure*}

We shall start our discussion from the resonant case when energy
levels in the both QDs are close to each other
$\varepsilon_1+U\simeq\varepsilon_2$. Fig.\ref{figure2}a,b
demonstrates filling numbers (localized charge) time evolution in
the first and second QDs ($n_1(t)$ and $n_2(t)$) for the different
values of Coulomb interaction. It is evident that Coulomb
interaction results in the increasing of the relaxation rate (grey
line on Fig.\ref{figure2}a,b) in comparison with the situation when
Coulomb interaction is absent (black line on Fig.\ref{figure2}a,b).
A particular value of Coulomb interaction exists in the system
($U/\gamma=4$ for a given set of parameters). When Coulomb
interaction is lower than $U/\gamma=4$, localized charge relaxation
occurs monotonically (Fig.\ref{figure2}a,b grey line). Otherwise one
can clearly see that relaxation process becomes non-monotonic and
reveals several typical time intervals with different values of
relaxation rates (black dashed line on Fig.\ref{figure2}a,b).
Obtained results strongly correspond to the localized charge
relaxation peculiarities in the system of coupled QDs when Coulomb
interaction is taken into account only within the second QD
\cite{Mantsevich}.

\begin{figure*} [t]
\includegraphics[width=110mm]{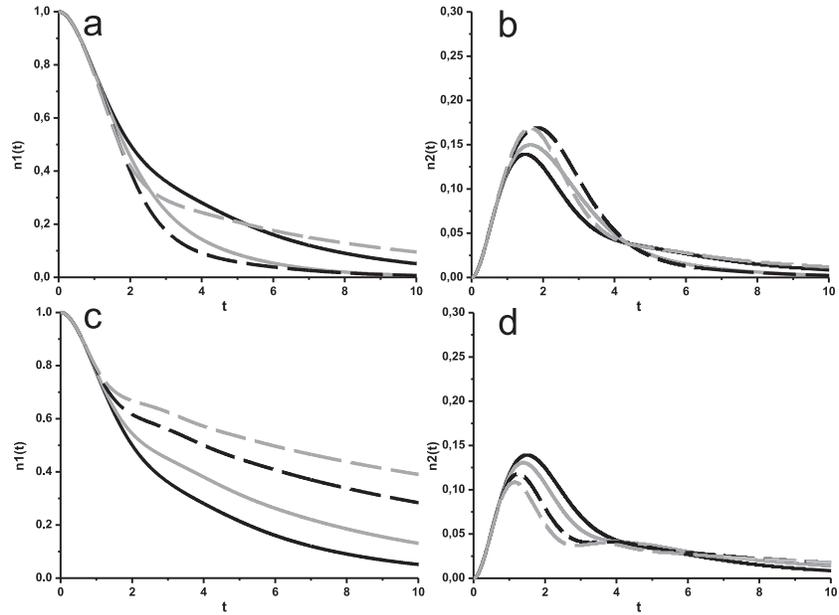}
\caption{Filling numbers evolution in the first a),c). $n_{1}(t)$
and second b),d). $n_{2}(t)$ QDs. Panels a). and b). correspond to
the positive detuning $(\varepsilon_1-\varepsilon_2)/\gamma=1.0$.
Panels c). and d). correspond to the negative detuning
$(\varepsilon_1-\varepsilon_2)/\gamma=-1.0$. Parameters
$T/\gamma=0.6$ and $\gamma=1.0$ are the same for all the figures.
$U=0$-black line, $U=0.5$-grey line, $U=1.5$-black dashed line,
$U=2.5$-grey dashed line.} \label{figure3}
\end{figure*}

Now let us analyze non-resonant case when difference between the
energy levels is about the values of parameters $T$ and $\gamma$.
We'll consider different signs of the initial detuning between
energy levels. If the detuning has positive value
($\varepsilon_1+U>\varepsilon_2$) for the small values of Coulomb
interaction ($U/\gamma=0.5$, $U/\gamma=1.0$) filling numbers
relaxation rate in the first QD increases in comparison with the
case when Coulomb interaction is absent (Fig.\ref{figure3}a). With
the increasing of Coulomb interaction value for
$\varepsilon_2>\varepsilon_1$ the condition
$|U(n_{1}(t)-1)|=\Delta\varepsilon$ is fulfilled. So, energy levels
detuning turns to zero at the particular time moment. Consequently,
resonant tunneling takes place and charge relaxation rate reaches
it's maximum value ($U/\gamma=1.5$ on the Fig.\ref{figure3}a). With
the further increasing of the Coulomb interaction detuning quickly
turns to zero and changes the sign.  It results in the decreasing of
relaxation rate ($U/\gamma=2.5$ on the Fig.3a). In the opposite case
of negative initial energy levels detuning
($\varepsilon_1+U<\varepsilon_2$) Coulomb interaction results in the
increasing of the detuning value and decreasing of the filling
numbers relaxation rate in the first QD (Fig.\ref{figure3}c).
Localized charge relaxation in this case reveals two time intervals
with different typical relaxation rate's scales. Relaxation rate in
the first time interval exceeds relaxation rate in the second one.

In the case of strong Coulomb interaction ($U/\gamma=4$ on the
Fig.\ref{figure2}c,d) one can distinguish three time intervals with
different typical relaxation rate's scales in the electron filling
number time evolution. These coincides with the results obtained for
the system of coupled QDs when Coulomb interaction  between
electrons is taken into account within the second QD
\cite{Mantsevich}.

\begin{figure*} [t]
\includegraphics[width=110mm]{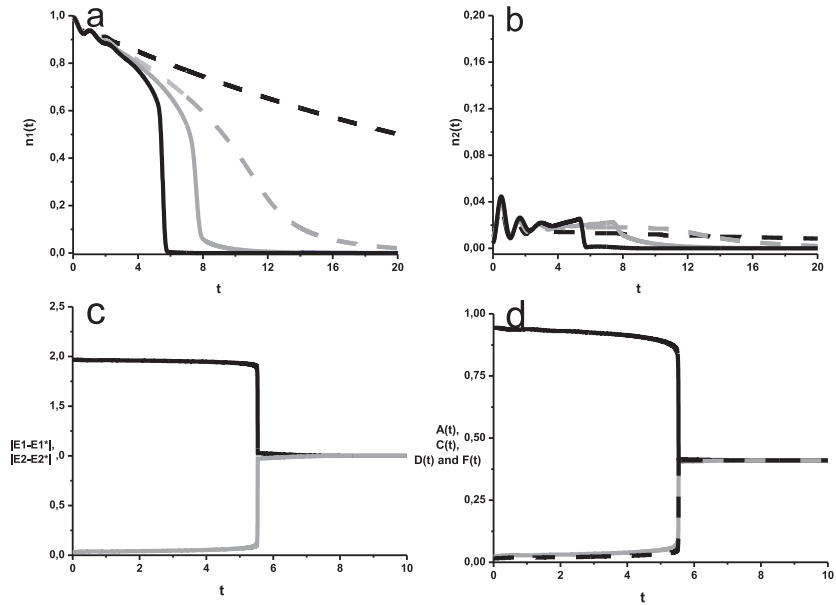}
\caption{a). Filling numbers evolution in the first QD $n_{1}(t)$.
$U=0$-dashed black line, $U=4$-dashed grey line, $U=6$-grey line,
$U=7$-black line.; b). Filling numbers evolution in the second QD
$n_{2}(t)$. $U=0$-dashed black line, $U=4$-dashed grey line,
$U=6$-grey line, $U=7$-black line; c). $|E1-E1^{*}|$-first and
$|E2-E2^{*}|$-second mode relaxation rates evolution for $U=7$; d).
First mode amplitude in the first QD time evolution $A(t)$-black
line; second mode amplitude in the first QD time evolution
$C(t)$-grey line; first and second modes in the second QD time
evolution $D(t)$ and $F(t)$-dashed black line for $U=7$. Parameters
$T/\gamma=0.8$, $\gamma=1.0$ and
$(\varepsilon_1-\varepsilon_2)/\gamma=6.0$ are the same for all the
figures.} \label{figure4}
\end{figure*}

Let us now focus on the most significant result obtained for the
system under investigation. It is clearly evident that when energy
levels detuning strongly exceeds Coulomb interaction, localized
charge relaxation occurs monotonically with a single typical value
of relaxation rate (black dashed line on Fig.\ref{figure4}a). We
revealed that when the Coulomb interaction value within the first QD
exceeds the values of tunneling transfer rates ($T$ and $\gamma$)
and becomes equal to the value of strong positive detuning between
energy levels ($\Delta\varepsilon>>T,\gamma$) in the QDs
($\Delta\varepsilon/\gamma\simeq U/\gamma$) or exceeds it, charge
relaxation occurs through two stable regimes which are characterized
by significantly different relaxation rate's values (black and grey
lines on Fig.\ref{figure4}).

When above-stated conditions are valid at a certain instant of time
charge relaxation rate changes discontinuously between two stable
values. This phenomenon in the localized charge evolution can be
called bifurcation.

With the increasing of Coulomb interaction bifurcation takes place
for the smaller time values (black and grey lines on
Fig.\ref{figure4}a). So one can tune the bifurcation moment by
changing the detuning value and the strength of Coulomb interaction.

We have not revealed bifurcations in the case when Coulomb
interaction was taken into account within the second QD
\cite{Mantsevich}. The following physical reason as an explanation
of this fact can be considered: Localized charge relaxation in the
first QD occurs only due to the tunneling processes to the second
QD. Relaxation from the second dot is possible due to the coupling
between the QDs and also to the transitions to continuous spectrum
states. So when Coulomb interaction is taken into account within the
first QD the charge relaxation demonstrates much more rough
behaviour due to the only one relaxation channel from this QD. It
results in the bifurcations formation.

For the detailed analysis of charge relaxation processes we shall
carefully examine power law exponents evolution, which determine
charge relaxation rates changing in each mode of the QDs
(Fig.\ref{figure4}c). Moreover we shall analyze time evolution of
preexponenial factors which reveal charge distribution among the
modes (Fig.\ref{figure4}d).

Power law exponents evolution is the same for the both QDs
(Fig.\ref{figure4}c). When the parameters values correspond to the
bifurcation regime in the charge evolution, relaxation rates of the
first and second modes demonstrate fast transitions between the two
stable values. First mode relaxation rate decreases and second mode
relaxation rate increases. After bifurcation both modes reveal
identical relaxation rates values (Fig.\ref{figure4}c).

Let us now analyze preexponential factors time evolution (mode's
amplitudes) in the presence of Coulomb interaction. In the second QD
time evolution of preexponential factors is determined by the same
law (expression \ref{p2}) (Fig.\ref{figure4}d). Time evolution of
the preexponential factors in the first QD significantly differs
(expression \ref{p1}). When the Coulomb interaction value becomes
equal to the detuning value, modes amplitudes demonstrate fast
switching between the two stable values. First mode amplitude in the
first QD rapidly decreases. Second mode amplitude in the first QD
and both mode's amplitudes in the second QD increase. After
bifurcation all mode's amplitudes admit equal values. It means that
Coulomb interaction leads to equal charge re-distribution among the
modes in the system of coupled QDs.

\section{Conclusion}

We have analyzed time evolution of localized charge in the system of
coupled QDs in the presence of Coulomb interaction between electrons
within a QD. We have found that Coulomb interaction strongly
modifies the relaxation rates and the character of localized charge
time evolution. It was shown that several time ranges with
considerably different relaxation rates arise in the system of the
two coupled QDs. We revealed that at a certain instant of time the
system switches rapidly between stable relaxation regimes in a wide
range of system parameters. This time moment can be tuned by
changing the detuning and Coulomb interaction values. We consider
this phenomenon to be applicable for active nano-electronics
elements creation based on the effect of fast switching between the
two stable states.

This work was partly supported by RFBR and Leading Scientific School
grants.


\pagebreak


\begin{thebibliography}{99}


\bibitem{Collier}
C.P. Collier, E.W. Wong, M. Belohradsky et. al., {\it Science},
\textbf{285}, (1999), 391.
\bibitem{Gittins}
D.I. Gittins et.al., {\it Nature}, \textbf{408}, (2000), 677.
\bibitem{Kastner}
M.A. Kastner, {\it Rev. Mod. Phys.}, \textbf{4}, (1992), 849.
\bibitem{Ashoori}
R. Ashoori, {\it Nature}, \textbf{379}, (1996), 413.
\bibitem{Oosterkamp}
T.H. Oosterkamp, T. Fujisawa, W.G. van der Wiel et.al., {\it
Nature}, \textbf{395}, (1998), 873.
\bibitem{Blick_0}
R.H. Blick, D. van der Weide, R.J. Haug et.al., {\it Phys.Rev
Lett.}, \textbf{81}, (1998), 689.
\bibitem{Bimberg}
D. Bimberg, M. Grundmann, N. Ledentsov, {\it Quantum Dot
Heterostructures} Wiley, New-York, (1999).
\bibitem{Livermore}
C. Livermore, C.H. Crouch, R.M. Westervelt et.al., {\it Science},
\textbf{274}, (1996), 1332.
\bibitem{Rotter}
I. Rotter, A.F. Sadreev, {\it Phys. Rev. E}, \textbf{71}, (2005),
036227.
\bibitem{Goldman}
V.J. Goldman, D.C. Tsui, J.E. Cunningham, {\it Phys. Rev. Lett.},
\textbf{58}, (1987), 1256.
\bibitem{Vamivakas}
A.N. Vamivakas, C.-Y. Lu, C. matthiesen et.al., {\it Nature
Letters}, \textbf{467}, (2010), 297.
\bibitem{Stinaff}
E.A. Stinaff, M. Scheibner, A.S. Bracker et.al., {\it Science},
\textbf{311}, (2006), 636.
\bibitem{Elzerman}
J.M. Elzerman, K.M. Weiss, J. Miguel-Sanchez et.al., {\it Phys. Rev.
Lett.}, \textbf{107}, (2011), 017401.
\bibitem{Peng}
J. Peng, G. Bester, {\it Phys. Rev. B}, \textbf{82}, (2010), 235314.
\bibitem{Munoz-Matutano}
G. Munoz-Matutano, M. Royo, J.I. Climente et.al., {\it Phys. Rev.
B}, \textbf{84}, (2011), 041308(R).
\bibitem{Angus}
S. J. Angus, A.J. Ferguson, A.S. Dzurak et. al., {\it Nano Lett.},
\textbf{7}, (2007), 2051.
\bibitem{Grove-Rasmussen}
K. Grove-Rasmussen, H. Jorgensen, T. Hayashi et. al., {\it Nano
Lett.}, \textbf{8}, (2008), 1055.
\bibitem{Moriyama}
S. Moriyama, D. Tsuya, E. Watanabe et. al., {\it Nano Lett.},
\textbf{9}, (2009), 2891.
\bibitem{Landauer}
R. Landauer, {\it Science}, \textbf{272}, (1996), 1914.
\bibitem{Mantsevich}
P.I. Arseyev, N.S. Maslova, V.N. Mantsevich, {\it arXiv: 1111.5776}.
\bibitem{Alexandrov}
A.S. Alexandrov, A.M. Bratkovsky, R.S. Williams, {\it Phys.Rev.B},
\textbf{67}, (2003), 075301.
\bibitem{Orellana}
P.A. Orellana, G.A. Lara, E.V. Anda, {\it Phys.Rev.B}, \textbf{65},
(2002), 155317.
\bibitem{Rack}
A.Rack, R. Wetzler, A. Wacker, E. Scholl, {\it Phys.Rev.B},
\textbf{66}, (2002), 165429.
\bibitem{Djuric}
I.Djuric, C.P. Search, {\it Phys.Rev.B}, \textbf{75}, (2007),
155307.
\bibitem{Coleman_1}
P.Coleman, {\it Phys.Rev.B}, \textbf{29}, (1984), 3035.
\bibitem{Coleman_2}
P.Coleman, {\it Phys.Rev.B}, \textbf{35}, (1987), 5072.
\bibitem{Wetzler}
R. Wetzler, C.M.A. Kapteyn, R. Heitz, A. Wacker, E. Sholl, D.
Bimberg, {\it Appl.Phys. Lett.}, \textbf{77}, (2000), 1671.
\bibitem{Hubbard}
J.Hubbard, {\it Proc.Roy.Soc.A}, \textbf{276}, (1963), 238.
\bibitem{Alexandrov}
A.S. Alexandrov, A.M. Bratkovsky, R.S. Williams, {\it Phys.Rev.B},
\textbf{67}, (2003), 075301.
\bibitem{Keldysh}
L.V. Keldysh, {\it Sov. Phys. JETP}, \textbf{20}, (1964), 1018.
\bibitem{Kikoin}
K.Kikoin, Y. Avishai, {\it Phys.Rev.Lett.}, \textbf{86}, (2001),
2090.
\bibitem{Contreras}
L.D. Contreras-Pulido, J. Splettstoesser, M. Governale, et.al., {\it
arXiv: 1111.4135}
\bibitem{Anderson}
P.W. Anderson, {\it Phys.Rev.}, \textbf{124}, (1961), 41.


\end{thebibliography}
\end{document}